\documentclass{iaus}
\usepackage{graphicx}

\def\apj{{ApJ}}			
\def\apjl{{ApJ}}

\def\aap{{A\&A}}

\def\pasj{{PASJ}}

\def\solphys{{Solar\ Phys.}}

\def\la{\mathrel{\hbox{\rlap{\hbox{\lower4pt\hbox{$\sim$}}}\hbox{$<$}}}}
\def\ga{\mathrel{\hbox{\rlap{\hbox{\lower4pt\hbox{$\sim$}}}\hbox{$>$}}}}

\def\rmit#1{{\it #1}}              
\def\rmit#1{{\rm #1}}              

\def\eg{\rmit{e.g.,}}              

\def\lesssim{\mathrel{\hbox{\rlap{\hbox{\lower4pt\hbox{$\sim$}}}\hbox{$<$}}}}
\def\gtrsim{\mathrel{\hbox{\rlap{\hbox{\lower4pt\hbox{$\sim$}}}\hbox{$>$}}}}

\def\level #1 #2#3#4{$#1 \: ^{#2} \mbox{#3} ^{#4}$}

\usepackage{natbib}

\title[Helicity of the solar magnetic field] 
{Helicity of the solar magnetic field}

\author[Sanjiv Kumar Tiwari]   
{Sanjiv Kumar Tiwari}

\affiliation{Udaipur Solar Observatory, Physical Research Laboratory,
Dewali, Bari Road,\\
Udaipur - 313 001, India.
\\ email: {\tt stiwari@prl.res.in}} 

\pubyear{2011}
\volume{273}  
\pagerange{1--8}
\jname{IAU-273, Physics of Sun and Star Spots}
\editors{D. P. Choudhary \& K. G. Strassmeier, eds.}
\begin{document}

\maketitle

\begin{abstract}
Helicity measures complexity in the field. Magnetic helicity is
given by a volume integral over the scalar product of magnetic field
{\bf B} and its vector potential {\bf A}. A direct computation of
magnetic helicity in the solar atmosphere is not possible due to
unavailability of the observations at different heights and also due
to non-uniqueness of {\bf A}. The force-free parameter $\alpha$ has
been used as a proxy of magnetic helicity for a long time. We have
clarified the physical meaning of $\alpha$ and its relationship with
the magnetic helicity. We have studied the effect of polarimetric
noise on estimation of various magnetic parameters. Fine structures
of sunspots in terms of vertical current ($J_z$) and $\alpha$ have
been examined. We have introduced the concept of signed shear angle
(SSA) for sunspots and established its importance for non force-free
fields. We find that there is no net current in sunspots even in
presence of a significant twist, showing consistency with their
fibril-bundle nature. The finding of existence of a lower limit of
SASSA for a given class of X-ray flare will be very useful for space
weather forecasting. A good correlation is found between the sign of
helicity in the sunspots and the chirality of the associated
chromospheric and coronal features. We find that a large number of
sunspots observed in the declining phase of solar cycle 23 do not
follow the hemispheric helicity rule whereas most of the sunspots
observed in the beginning of new solar cycle 24 do follow. This
indicates a long term behaviour of the hemispheric helicity patterns
in the Sun. The above sums up my PhD thesis.
\end{abstract}

\vspace*{-0.3 cm}
\section{Introduction}
Magnetic helicity is a physical quantity that measures the degree of
linkage and twistedness in the magnetic field lines \citep{moff78}.
It is given as
\begin{equation}
H_m = \int {\mathbf A\cdot \bf B} ~dV
\end{equation}
The term magnetic helicity was introduced by \cite{els56} and many
of its important characteristics were studied by
\cite{wolt58,tayl74,berg84} etc.

The handedness associated with the field is defined by `chirality'.
Helicity is closely related to chirality. If the twist on the
surface is clockwise, the chirality is negative and the field bears
dextral chirality. The sunspot twist direction is decided by the
curvature of sunspot whirls \citep{mart98,tiw09e}. If the twist is
counterclockwise (when we go from sunspot center towards outside),
the chirality is sinistral and sign of helicity is positive. Reverse
is true for the dextral chirality. These definitions of chirality
have been used to study the hemispheric patterns of the active
regions \citep{tiw08,tiw10b}.

One of the main motivations of the thesis \citep{tiw09e} was to use
the helicity or related parameters to help in predicting the
severity of the solar flares. If done so, this would contribute in
improving the space-weather forecasting. We have found the parameter
signed shear angle (SSA) to be very useful in this context \citep{tiw10a}.\\
Some of the important results of my Ph.D. thesis are summarized very
briefly in the following sections.

\section{Estimating magnetic parameters}
{\underline{\it Physical meaning of $\alpha$}}. We arrive at the
following depiction of $\alpha$ (for details, please see Appendix A
of \cite{tiw09a}):
\begin{equation}\label{alfinal}
{\alpha = 2 ~\frac{ d \phi}{dz}}\\
\end{equation}
From Equation \ref{alfinal}, it is clear that the $\alpha$  gives
twice the degree of twist per unit axial length. If we take one
complete rotation of flux tube i.e., $\phi= 2\pi$, and loop length
\textsc{$\lambda \approx 10^{9}$} meters, then
\begin{equation}\label{}
    {\alpha = \frac{2\times 2\pi}{\lambda}}
\end{equation}
comes out to be of the order of $10^{-8}$ per meter.

{\underline{\it Correlation between sign of H$_m$ and that of
$\alpha$}}. Vector potential in terms of scalar potential $\phi$ can
be expressed as (for details, please see Appendix B of
\cite{tiw09a})
\begin{equation}\label{}
    \bf A = \bf B \alpha^{-1} + \nabla \phi
\end{equation}
which is valid only for constant $\alpha$. Using this relation in
Equation 1, we get magnetic helicity as
\begin{eqnarray}\label{hm1}
\nonumber  H_{m} &=& \int ({\bf B} \alpha^{-1} + \nabla \phi) \cdot {\bf B} ~ dV \\
        &=& \int B^2 \alpha^{-1} dV + \int ({\bf B} \cdot \nabla) \phi~dV \sim(\int (\phi\ {\bf B})\cdot {\bf n}~ dS)
\end{eqnarray}
showing that the force free parameter $\alpha$ has the same sign as
that of the magnetic helicity iff $\bf n \cdot B = 0$ i.e., no field
lines cross the boundary, which is not the case with the Sun.

{\underline{\it A direct method for calculating global $\alpha$}}.
We prefer to use the second moment of minimization \citep{tiw09a}
leading to the following expression:
\begin{equation}\label{}
\alpha_{g}=\frac{\sum(\frac{\partial B_y}{\partial x} -
\frac{\partial B_x}{\partial y})B_z}{\sum B_z^2}.
\end{equation}
This formula gives a single global value of $\alpha$ in a sunspot
and is the similar to $\alpha_{av}^{(2)}$ of \cite{hagi04}. We do
not use direct mean ($0^{th}$ order moment) as it leads to
singularities at neutral lines where B$_z \sim 0$. First order
moment will also lead to singularities when flux is balanced.

{\underline{\it Estimating the effect of polarimetric noise in the
measurement of field parameters}}. Using the analytical bipole
method \citep{low82}, non-potential force-free field components
B$_x$, B$_y$ \& B$_z$ in a plane have been generated. We calculate
the synthetic Stokes profiles for each B, $\gamma$ and $\xi$ in a
grid of 100 x 100 pixels, using the He-Line Information Extractor
``HELIX'' code \citep{lagg04}. We add random noise of 0.5 \% of the
continuum intensity I$_{c}$ \citep{ichi08} to the polarimetric
profiles as observed in SOT/SP aboard Hinode. In addition, we also
study the effect of adding a noise of 2.0\% level to Stokes profiles
as a worst case scenario. We add 100 realizations of the noise of
the orders mentioned above to each pixel and invert the
corresponding 100 noisy profiles using the ``HELIX'' code. The
effect of polarimetric noise in the derivation of vector fields and
other parameters such as $\alpha_g$ and magnetic energy is found to
be very small. We have done similar investigations as a second step
using real data \citep{gosain10}. As a third step we plan to use MHD
simulated data to check the inversion codes including the effect of
optical depth corrugation.

\section{Global twist of sunspot magnetic fields}
{\underline{\it Introduction of signed shear angle (SSA)}}. To
emphasize the sign of shear angle we introduce the signed shear
angle (SSA) for the sunspots as follows: choose an initial reference
azimuth for a current-free field (obtained from the observed line of
sight field). Then move to the observed field azimuth from the
reference azimuth through an acute angle. If this rotation is
counter-clockwise, then assign a positive sign for the SSA. A
negative sign is given for clockwise rotation. This sign convention
will be consistent with the sense of azimuthal field produced by a
vertical current. This sign convention is also consistent with the
sense of chirality \citep{tiw09b}. The SSA is computed from the
following formula \citep{tiw10a}:
\begin{equation}\label{}
SSA = \tan^{-1} (\frac{B_{yo} B_{xp} - B_{yp} B_{xo}}{B_{xo} B_{xp} + B_{yo} B_{yp}})
\end{equation}
where $B_{xo}, B_{yo}$ and $B_{xp}, B_{yp}$ are observed and
potential transverse components of sunspot magnetic fields
respectively. A spatial average of the SSA (SASSA) gives the global
twist of sunspot magnetic fields at observed height irrespective of
the force-free nature of the field and shape of sunspots
\citep{venk09}.

{\underline{\it Fine structures in terms of J$_z$ and $\alpha$}}.
Local J$_z$ and $\alpha$ patches of opposite signs are present in
the umbra of each sunspot. The amplitude of the spatial variation of
local $\alpha$ in the umbra is typically of the order of the global
$\alpha$ of the sunspot. We find that the local $\alpha$ is
distributed as alternately positive and negative filaments in the
penumbra. The amplitude of azimuthal variation of the local $\alpha$
in the penumbra is approximately an order of magnitude larger than
that in the umbra. The contributions of the local positive and
negative currents and $\alpha$ in the penumbra cancel each other
giving almost no contribution for their global values for whole
sunspot. The data sets used in the analysis are taken from ASP/DLSP
and Hinode (SOT/SP). See for details: \cite{tiw09b}. Most of the
data sets we studied are observed during the declining minimum phase
of solar cycle 23. All except 5, out of 43 sunspots observed, follow
the reverse twist hemispheric rule, while 5 follow the conventional
helicity rule. Also, $\alpha_g$  has same sign as the SASSA and
therefore the same sign of the photospheric chirality of the
sunspots, but the magnitudes of SASSA and $\alpha_{g}$ are not well
correlated. This lack of correlation could be due to a variety of
reasons: (a) departure from the force-free nature (b) even for the
force-free fields, $\alpha$ is the gradient of twist variation
whereas SASSA is purely an angle. The missing link is the scale
length of variation of twist.

\section{Net current in sunspots}
{\underline{\it Expression for net current}}. We consider a long
straight flux bundle surrounded by a region of ``field free'' plasma
following \cite{parker96}. \cite{parker96} assumed azimuthal
symmetry as well as zero radial component $B_r$, of the magnetic
field. For realistic sunspot fields, we have already seen the
ubiquitous fine structure of the radial magnetic field. Hence, we
need to relax both these assumptions.

The vertical component of the electric current density consists of
two terms, viz. $-\frac{1}{\mu_0 r}\frac{\partial
B_r}{\partial\psi}$ ~and~ $\frac{1}{\mu_0 r}\frac{\partial (r
B_{\psi})}{\partial r}$. We will call the first term as the ``pleat
current density", $j_p$ and the second term as the ``twist current
density", $j_t$. The total current $I_z$ within a distance $\varpi$
from the center is then given by
\begin{equation}\label{}
 I_z(\varpi) = \int_0^{2\pi} d\psi \int_0^\varpi rdr(j_p + j_t)
\end{equation}
The $\psi$ integral over $j_p$ vanishes, while the second term
yields
\begin{equation}\label{netcurr}
I_z(\varpi) = \frac{\varpi}{\mu_0} \int_0^{2\pi} d\psi B_\psi(\varpi,\psi)
\end{equation}
which gives the net currents within a circular region of radius
$\varpi$. The transverse vector can be expressed in cylindrical
geometry as
\begin{eqnarray}\label{}
B_r = \frac{1}{r}(xB_x + yB_y)\\
B_\psi = \frac{1}{r}(-yB_x + xB_y)
\end{eqnarray}\label{}
The azimuthal field $B_\psi$ is then used in Equation \ref{netcurr}
for obtaining the value for the total vertical current within a
radius $\varpi$.

{\underline{\it No net current: an evidence for fibril bundle nature
of sunspot magnetic field?}} As expected from the trend in Figure 3
of \cite{venk09}, the net current shows evidence for a rapid decline
after reaching a maximum. Similar trends were seen in other
sunspots. This can be interpreted as evidence for the neutralization
of the net current. Table 1 of \cite{venk09} shows the summary of
results for all the sunspots analyzed. Along with the power law
index $\delta$ of $B_{\psi}$ decrease, we have also shown the
average deviation of the azimuth from the radial direction (``twist
angle = $tan^{-1}(B_{\psi}/B_r)$"), as well as the SASSA. The
average deviation of the azimuth is well correlated with the SASSA
for nearly circular sunspots, but is not correlated with SASSA for
more irregularly shaped sunspots. Thus, SASSA is a more general
measure of the global twist of sunspots, irrespective of their
shape.

As is well known for astrophysical plasmas, that the plasma distorts
the magnetic field and the curl of this distorted field produces a
current by Ampere's law \citep{parker79}. Parker's (1996)
expectation of net zero current in a sunspot was basically motivated
by the concept of a fibril structure for the sunspot field. However,
he also did not rule out the possibility of vanishing net current
for a monolithic field where the azimuthal component of the vector
field in a cylindrical geometry declines faster than 1/$\varpi$.
While it is difficult to detect fibrils using the Zeeman effect,
notwithstanding the superior resolution of SOT on {\it Hinode}, the
stability and accuracy of the measurements have allowed us to detect
the faster than 1/$\varpi$ decline of the azimuthal component of the
magnetic field, which in turn can be construed as evidence for the
confinement of the sunspot field by the external plasma. The
resulting pattern of curl {\bf B} appears as a sharp decline in the
net current at the sunspot boundary. Although the existence of a
global twist in the absence of a net current is possible for a
monolithic sunspot field \citep{baty00}, a fibril model of the
sunspot field can accommodate a global twist even without a net
current \citep{parker96}. A sunspot, made up of a bundle of
magnetically isolated current free fibrils, can be given an overall
torsion without inducing a global current. For details and more
discussions please see \cite{venk09,tiw09e,tiw10c}.

\section{Relationship between the SASSA of active regions and
associated GOES X-ray flux} We find an upper limit of peak X-ray
flux for a given value of SASSA can be given for different classes
of X-ray flares. Figures 5(a) and 5(b) of \cite{tiw10a} represent
scatter plots between the peak GOES X-ray flux and interpolated
SASSA and mean weighted shear angle (MWSA: \cite{wang92}) values for
that time, respectively. The cubic spline interpolation of the
sample of the SASSA and the MWSA values has been done to get the
SASSA and MWSA exactly at the time of peak flux of the X-ray flare.
For details kindly see \cite{tiw10a}. We find that the SASSA, apart
from its helicity sign related studies, can also be used to predict
the severity of the solar flares. However to establish these lower
limits of SASSA for different classes of X-ray flares, we need more
cases to study. The SASSA already gives a good indication of its
utility from the present four case studies using 115 vector
magnetograms from Hinode (SOT/SP). Once the vector magnetograms are
routinely available with higher cadence, the lower limit of SASSA
for each class of X-ray flare can be established by calculating the
SASSA in a series of vector magnetograms. This will provide the
inputs to space weather models. Also, SASSA has shown a good
correlation with the free magnetic energy computed by \cite{jing10}.

The other non-potentiality parameter MWSA studied in \cite{tiw10a}
does show a similar trend as that of the SASSA. The magnitudes of
MWSA, however, do not show consistent threshold values as related
with the peak GOES X-ray flux of different classes of solar flares.
One possible reason for this behavior may be explained as follows:
The MWSA weights the strong transverse fields e.g., penumbral
fields. From the recent studies
\citep{su09,tiw09b,tiw09e,venk09,venk10} it is clear that the
penumbral field contains complicated structures with opposite signs
of vertical current and vertical component of the magnetic tension
forces. Although the amplitudes of the magnetic parameters are found
high in the penumbra, they do not contribute to their global values
because they contain opposite signs, which cancel out in the
averaging process \citep{tiw09e,tiw09b}. On the other hand, the MWSA
adds those high values of shear and produces a pedestal that might
mask any the relation between the more relevant global
non-potentiality and the peak X-ray flux. Whereas the SASSA perhaps
gives more relevant value of the shear after cancelation of the
penumbral contribution.

\section{Solar cycle dependence}
{\underline{\it Helicity hemispheric rule}}. We compare the
behaviour of magnetic helicity sign of AR's observed in the
beginning of 24$^{th}$ solar cycle with some AR's observed in the
declining phase of 23$^{rd}$ solar cycle. We find that the majority
of active regions in the beginning of solar cycle 24 do follow the
hemispheric helicity rule whereas those observed in the declining
phase of solar cycle 23 do not (Tiwari, 2009; Tiwari et al., 2010c,
in prep.).

{\underline{\it Sign of magnetic helicity at different heights in
the solar atmosphere}}. A good correlation has been found among the
sign of helicity in the associated features observed at
photospheric, chromospheric and coronal heights without solar cycle
dependence (Tiwari et al. 2008; Tiwari, 2009; Tiwari et al. 2010b;
Tiwari et al., 2010c, in prep.).

\section{Conclusions}
The magnetic field parameters can be derived very accurately using
the recent data available (\eg\ from {\it Hinode} (SOT/SP)) and
advanced inversion codes. The SASSA is the best measure of the
global magnetic twist of sunspot magnetic fields at observed height,
irrespective of the force-free nature and the shape of sunspots. The
sunspots with significant twist and no net currents show consistency
with the fibril bundle nature of the sunspots. The study of
evolution of SASSA of sunspots showed threshold values for different
classes of X-ray flares. This is an important discovery which was
being sought after for many decades. The magnetic helicity sign of
active regions studied, has good correlation with the sign of
chirality of associated features observed at chromospheric and
coronal heights. The majority of sunspots observed in the declining
phase of solar cycle 23 follow a reverse hemispheric helicity rule,
whereas most of the AR's emerged in the beginning of solar cycle 24
follow the conventional helicity rule. This result indicates that
revisiting the hemispheric helicity rule using data sets of several
years is required.

\acknowledgments The presentation of this paper in the IAU Symposium
273 was possible due to  partial support from the National Science
Foundation grant numbers ATM 0548260, AST 0968672 and NASA - Living
With a Star grant number 09-LWSTRT09-0039. I am grateful to
Professor P. Venkatakrishnan for his invaluable and patient guidance
during my PhD thesis. Hinode is a Japanese mission developed and
launched by ISAS/JAXA, with NAOJ as domestic partner and NASA and
STFC (UK) as international partners. It is operated by these
agencies in co-operation with ESA and NSC (Norway).

\end{document}